\documentclass[aps,prl,showpacs,twocolumn,superscriptaddress,floatfix]{revtex4-2}
\usepackage[utf8]{inputenc}
\newcommand{\figurescale}{1}
\usepackage{verbatim}
\usepackage{amsmath}
\usepackage{mathtools}

\DeclarePairedDelimiterX\braket[2]{\langle}{\rangle}{#1 \delimsize\vert #2}

\usepackage[pdftex]{graphicx}
\usepackage[separate-uncertainty=true]{siunitx}
\DeclareSIUnit{\rpm}{rpm}
\usepackage[colorlinks=true,urlcolor=blue,linkcolor=blue,citecolor=blue]{hyperref}
\usepackage[usenames,dvipsnames]{xcolor}
\usepackage[T1]{fontenc}
\usepackage{placeins}
\usepackage{nicefrac}
\usepackage{braket}
\usepackage{pdfcomment}
\usepackage{xcolor}
\usepackage{braket}
\usepackage[normalem]{ulem}

\usepackage{lineno}

\begin{document}

\title{Trions in MoS$_2$ are quantum superpositions of \textit{intra}- and \textit{inter}valley spin states}
%
\author{J.~Klein}\email{jpklein@mit.edu}
\affiliation{Walter Schottky Institut and Physik Department, Technische Universit\"at M\"unchen, Am Coulombwall 4, 85748 Garching, Germany}
\affiliation{Department of Materials Science and Engineering, Massachusetts Institute of Technology, Cambridge, Massachusetts 02139, USA}
\author{M.~Florian}
\affiliation{Institut für Theoretische Physik, Universität Bremen, P.O. Box 330 440, 28334 Bremen, Germany}
\affiliation{Department of Electrical Engineering and Computer Science, University of Michigan, Ann Arbor, MI, USA}
\author{A.~H\"otger}
\affiliation{Walter Schottky Institut and Physik Department, Technische Universit\"at M\"unchen, Am Coulombwall 4, 85748 Garching, Germany}
\author{A.~Steinhoff}
\affiliation{Institut für Theoretische Physik, Universität Bremen, P.O. Box 330 440, 28334 Bremen, Germany}
\author{A.~Delhomme}
\affiliation{Universit\'e Grenoble Alpes, INSA Toulouse, Univ. Toulouse Paul Sabatier, EMFL, CNRS, LNCMI, 38000 Grenoble, France.}
\author{T.~Taniguchi}
\affiliation{International Center for Materials Nanoarchitectonics,
National Institute for Materials Science,  1-1 Namiki, Tsukuba 305-0044, Japan
}
\author{K.~Watanabe}
\affiliation{Research Center for Functional Materials,
National Institute for Materials Science, 1-1 Namiki, Tsukuba 305-0044, Japan
}
\author{F.~Jahnke}
\affiliation{Institut für Theoretische Physik, Universität Bremen, P.O. Box 330 440, 28334 Bremen, Germany}
\author{A.~W.~Holleitner}
\affiliation{Walter Schottky Institut and Physik Department, Technische Universit\"at M\"unchen, Am Coulombwall 4, 85748 Garching, Germany}
\author{M.~Potemski}
\affiliation{Universit\'e Grenoble Alpes, INSA Toulouse, Univ. Toulouse Paul Sabatier, EMFL, CNRS, LNCMI, 38000 Grenoble, France.}
\author{C.~Faugeras}
\affiliation{Universit\'e Grenoble Alpes, INSA Toulouse, Univ. Toulouse Paul Sabatier, EMFL, CNRS, LNCMI, 38000 Grenoble, France.}
\author{A.~V.~Stier}\email{andreas.stier@wsi.tum.de}
\affiliation{Walter Schottky Institut and Physik Department, Technische Universit\"at M\"unchen, Am Coulombwall 4, 85748 Garching, Germany}
\author{J.~J.~Finley}\email{finley@wsi.tum.de}
\affiliation{Walter Schottky Institut and Physik Department, Technische Universit\"at M\"unchen, Am Coulombwall 4, 85748 Garching, Germany}
%

%
\date{\today}
%
%
\begin{abstract}
We report magneto-photoluminescence spectroscopy of gated  MoS$_2$ monolayers in high magnetic fields to 28 T. At $B = \SI{0}{\tesla}$ and electron density $n_s\sim 10^{12} \SI{}{\per\centi\meter\squared}$, we observe three trion resonances that cannot be explained within a single-particle picture. Employing ab initio calculations that take into account three-particle correlation effects as well as local and non-local electron-hole exchange interaction, we identify those features as quantum superpositions of \textit{inter}- and \textit{intra}valley spin states. We experimentally investigate the mixed character of the trion wave function via the filling factor dependent valley Zeeman shift in positive and negative magnetic fields. Our results highlight the importance of exchange interactions for exciton physics in monolayer MoS$_2$ and provide new insights into the microscopic understanding of trion physics in 2D multi-valley semiconductors for low excess carrier densities.
\end{abstract}
%
%
\maketitle
%
%

Early experiments on quasi two-dimensional CdTe~\cite{Kheng.1993} and GaAs~\cite{Finkelstein.1995,Buhmann.1995,BarAd.1992} quantum wells allowed the first observation of charged exciton complexes owing to an increase in the exciton binding energy arising from confinement effects. Atomically thin transition metal dichalcogenides (TMDCs) of formula MX$_2$ where M = Mo, W and X = S, Se or Te are excellent model systems for studying excitonic physics in two-dimensional (2D) systems, due to enhanced quantum confinement and weak dielectric screening~\cite{Chernikov.2014,Xu.2014,Wang.2018,Stier.2016,Stier.2018}. The inherent 2D nature, broken spatial inversion symmetry and strong spin-valley optical selection rules~\cite{Xiao.2012} open up a plethora of possibilities for the controlled study of exciton physics in the presence of free carriers~\cite{Mak.2013,Ross.2013,Chernikov.2015,Sidler.2016,Barbone.2018}. Combined with the ability to integrate monolayers into functional devices, such experiments can be performed with full control of the local charge density. Unlike quasi-2D quantum well systems, valley dichroism and strong spin-orbit splitting in the conduction and valence bands promotes the formation of dipole allowed trion complexes having singlet and triplet spin structure~\cite{Yu.2014,Jones.2015,Plechinger.2016,Courtade.2017,Drppel.2017,Wang.2017,Arora.2020}. In WSe$_2$ and WS$_2$, the lowest energy exciton is spin forbidden due to a dark band alignment arising from the large conduction band spin-orbit splitting of $\Delta_{CB} \sim \SI{30}{\milli\electronvolt}$~\cite{Kormnyos.2015,Wang.2016,Zhou.2017,Zhang.2017,Molas.2017}. This is in stark contrast to MoSe$_2$~\cite{Ross.2013}, a material that is considered to be optically bright since the lowest exciton transition is spin allowed~\cite{Kormnyos.2015,Zhou.2017,Lu.2019,Robert.2020,Arora.2020}.

\begin{figure}
	\scalebox{\figurescale}{\includegraphics[width=1\linewidth]{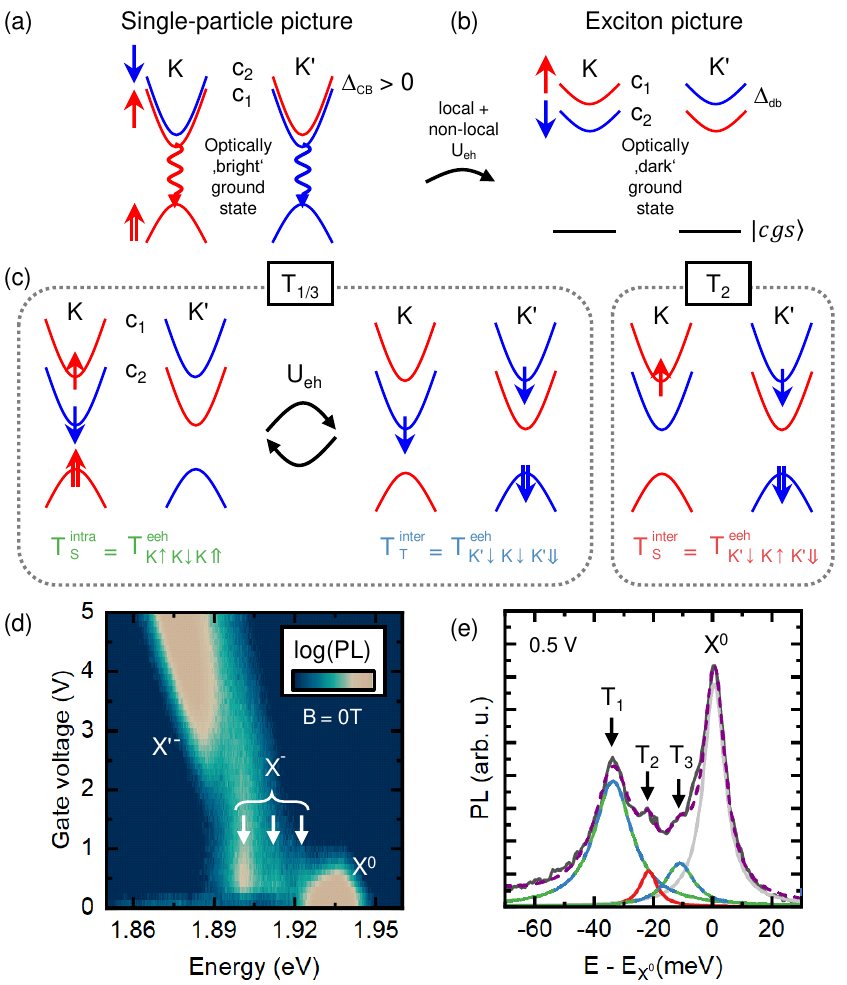}}
	\renewcommand{\figurename}{FIG.|}
	\caption{\label{fig1}
		%
		(a) Monolayer MoS$_2$ is optically bright in a single-particle picture with $\Delta_{CB}>0$.
		(b) Local exchange interaction $U_{eh}$ re-orders exciton spin configurations resulting in the lowest transition to be optically dark. This is schematically depicted as a flip of conduction bands.
		(c) Trion fine structure in monolayer MoS$_{2}$:
		Three bright configurations are given by the \textit{intra}valley singlet $T^{\text{intra}}_{S} = T^{eeh}_{K\uparrow K\downarrow K\Uparrow}$ (left) and \textit{inter}valley triplet trion $T^{\text{inter}}_{T} = T^{eeh}_{K^{\prime}\downarrow K\downarrow K^{\prime}\Downarrow}$ (center), which are coupled by non-local electron-hole exchange interaction $U_{eh}$, as well as the \textit{inter}valley singlet trion $T^{\text{inter}}_{S} = T^{eeh}_{K^{\prime}\downarrow K\uparrow K^{\prime}\Downarrow}$ (right). The trion amplitude denoted by $T^{e_1e_2h_3}_{\mathbf k_1 \mathbf k_2 \mathbf k_3}=\braket{{(a_{\mathbf{k_1+k_2+k_3}}^{e_2}})^\dagger a_{\mathbf{k}_3}^{h_3} a_{\mathbf{k}_2}^{e_2} a_{\mathbf{k}_1}^{e_1}}$ is a four-operator expectation value and describes the correlated process of annihilating two electrons and one hole, leaving behind an electron with momentum $\mathbf Q=\mathbf{k_1+k_2+k_3}$ in the conduction band, and is linked to the optical response of an electron trion. We adopt the band ordering from the exciton picture.
		(d) False color plot of the gate voltage dependent $\SI{5}{\kelvin}$ PL showing the neutral exciton $X^{0}$ and a fine structure of negatively charged trions $X^{-}$ (highlighted with arrows) and a many-body state $X^{\prime-}$ at $\SI{2}{\volt}$ (densities of $n_s > 4 \cdot 10^{12} \SI{}{\per\centi\meter\squared}$)
		(e) Fitted PL spectrum at $\SI{0.5}{\volt}$ ($n_s \sim 10^{12}\SI{}{\per\centi\meter\squared}$) showing three distinct trion resonances $T_1$, $T_2$ and $T_3$.
		}
\end{figure}

The situation for MoS$_2$ is, however, more delicate. Initial experiments showed an increase of exciton luminescence intensity with temperature~\cite{Zhang.2015,Arora.2015}, which is a clear signature for an optically bright material, supporting early theoretical works~\cite{Liu.2013,Kormnyos.2015}. However, unlike optically bright MoSe$_2$, monolayer MoS$_2$ shows a large degree of valley polarization that is typically found in the optically dark materials, like WSe$_2$ and WS$_2$. Importantly, recent experiments on magnetic brightening of dark excitons unequivocally showed an optically dark alignment with a splitting between 1s states of bright and dark excitons of $\Delta_{db} \sim \SI{14}{\milli\electronvolt}$~\cite{Robert.2020}, which is consistent with more recent theoretical work~\cite{Deilmann.2017,Torche.2019,Deilmann.2020,Bieniek.2020}. This value reflects both the SOC in the conduction band and the difference in the effective mass of the two subbands which leads to this inversion of ground state between single-particle and excitonic picture. As such, the single-particle conduction band structure of MoS$_2$ is non-trivial due to the small spin splitting and it is altered by interactions in the exciton picture (see Fig.~\ref{fig1}(a) and (b)). Furthermore, a hallmark of optically dark materials is the appearance of a rich fine structure of excitonic complexes~\cite{Qiu.2015,steinhoff_biexciton_2018} as recently observed experimentally~\cite{Arora.2020,jadczak.2021,Grzeszczyk.2020}. However, these initial experiments lacked a gate to control the local charge carrier density and a fully developed microscopic understanding of the observed spectra. This motivates detailed charge-carrier-dependent investigations of the trion fine structure in monolayer MoS$_2$.


In this Letter, we show that the local and non-local exchange interactions ($U_{eh}$) determine the band alignment and corresponding trion fine structure in monolayer MoS$_2$. We further calculate how the mixing of unperturbed \textit{inter}/\textit{intra}valley trion states is driven by the exchange interactions and therefore modify the binding energies and wave function contributions of the observed trion features. This has strong implications for the interpretation of magneto-optical data in many works on 2D multi-valley semiconductors ~\cite{Aivazian.2015,Srivastava.2015,MacNeill.2015,Li.2014,Wang.2016,Stier.2016,Stier.2016b,Mitioglu.2016,Cadiz.2017,Smoleski.2019,Lyons.2019,Goryca.2019,Liu.2020,jadczak.2021,Wang.2020}. 

The band structure of monolayer MoS$_2$ deviates from other semiconducting TMDCs due to the small conduction band splitting in the single-particle picture. While the ground state is optically bright (see Fig.~\ref{fig1}(a)), small many-body effects can markedly alter the band structure ~\cite{Klein.2021,Roch.2019}. Local electron-hole exchange interaction leads to an overall blue shift of like-spin excitons that re-orders excitonic transitions and results in an optically dark ground state (see Fig.~\ref{fig1}(b) and Supplementary Material). In addition, non-local electron-hole exchange mixes excitons in the $K$ and $K^{\prime}$ valleys, resulting in a non-analytic light-like exciton dispersion for like-spin excitons. A detailed discussion for monolayer MoS$_2$ is given in Ref.~\cite{Qiu.2015}.



Furthermore, these interactions result in three trion configurations (Figure~\ref{fig1}(c)), which are quantum superpositions of the eigenstates of the three-body problem without electron-hole exchange interaction: an \textit{intra}valley singlet $T^{\text{intra}}_{S} = T^{eeh}_{K\uparrow K\downarrow K\Uparrow}$ and an \textit{inter}valley triplet trion $T^{\text{inter}}_{T} = T^{eeh}_{K^{\prime}\downarrow K\downarrow K^{\prime}\Downarrow}$ that are coupled due to non-local electron-hole exchange interaction $U_{eh}$ as well as an \textit{inter}valley singlet trion $T^{\text{inter}}_{S} = T^{eeh}_{K^{\prime}\downarrow K\uparrow K^{\prime}\Downarrow}$ where both electrons are located in the upper conduction bands $c_1$.
Under applied magnetic fields in the range $B = \pm \SI{28}{\tesla}$, we observe quantum oscillations in the intensity of the dominant trion emission line and observe that the polarity of the magnetic field switches the PL between \textit{intra}valley singlet and \textit{inter}valley triplet character due to time-reversal symmetry breaking and Landau level quantization. From our data we draw two main conclusions: (i) Trions observed in optical spectra have mixed wave function character, thus representing quantum superpositions with contributions from \textit{intra}valley and \textit{inter}valley singlet and triplet trions and (ii) the non-uniform and tunable trion $g$-factor results from the decay of the trion into a photon and a free electron sequentially occupying Landau levels in the $K$ and $K^{\prime}$ conduction bands.

We probe trion emission from an exfoliated monolayer MoS$_2$ in a commonly used gate-tunable van der Waals device structure (See Supplemental Material for additional information on the device)~\cite{Barbone.2018}.
The monolayer MoS$_2$ is fully encapsulated between thin layers of hBN ($\sim \SI{10}{\nano\meter}$) to reduce inhomogeneous linewidth broadening~\cite{Wierzbowski.2017,Cadiz.2017}. We apply a bias voltage between MoS$_2$ and a thin graphite bottom gate to control the carrier concentration $n_s$ in the device~\cite{Klein.2021}. 
Low-temperature $\SI{5}{\kelvin}$ magneto-photoluminescence measurements were performed using unpolarized CW laser excitation at $E = \SI{2.41}{\electronvolt}$ and $\sigma^-$ circularly polarized detection. A typical false color plot of gate voltage dependent PL obtained from our device is presented in Fig.~\ref{fig1}(d) and Fig. SM1. At zero carrier density, the PL is dominated by the neutral exciton $X^{0}$, as expected. When the electron density is increased, the oscillator strength shifts away from $X^0$ due to the formation of negatively charged excitons before spectral weight transitions to a many-body state $X^{\prime-}$ at $>\SI{2}{\volt}$ (densities $n_s > 4 \cdot 10^{12} \SI{}{\per\centi\meter\squared}$)~\cite{Sidler.2016,Klein.2021,Roch.2019}. Interestingly, besides the neutral exciton $X^{0}$ close to $n_s = 0$, three distinct trion resonances, $T_1$, $T_2$ and $T_3$ are clearly visible for a gate voltage of $\SI{0.5}{\volt}$ (electron density of $n_s \sim 1 \cdot10^{12} \SI{}{\per\centi\meter\squared}$) further highlighted by the fit in Fig.~\ref{fig1}(e). The voltage dependent oscillator strength due to band population of the individual trions is non-trivial due to the delicate interplay between single-particle and exciton picture in which the conduction band configuration re-orders due to non-local exchange interaction.

\begin{figure}
	\scalebox{\figurescale}{\includegraphics[width=1\linewidth]{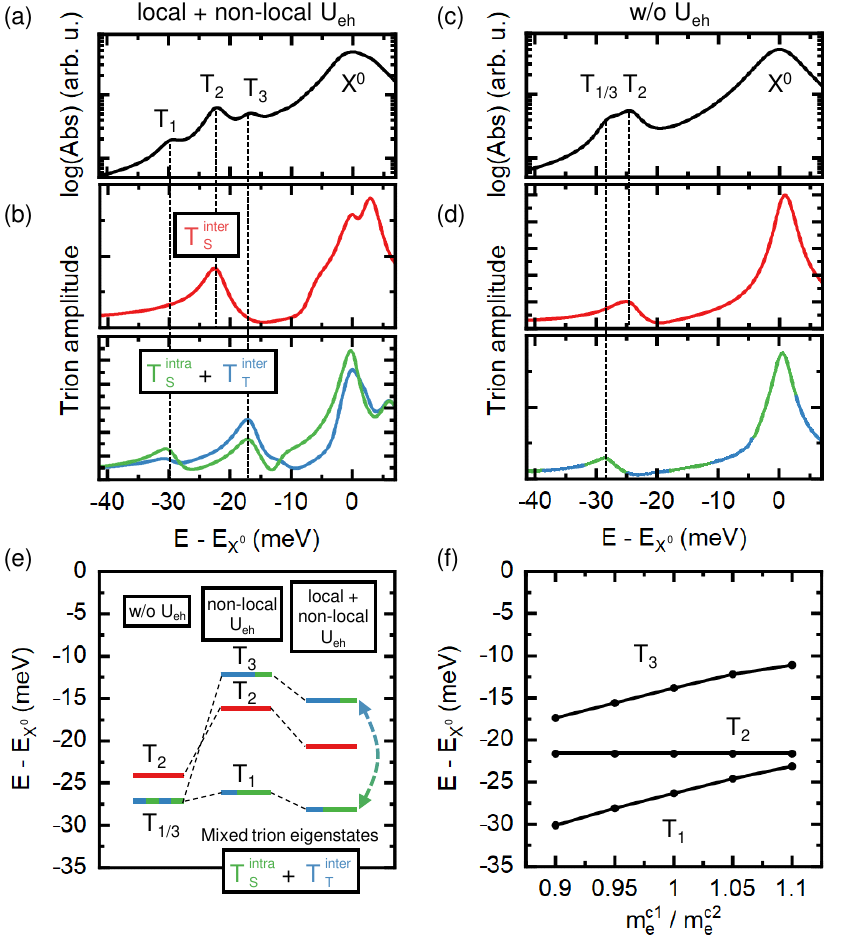}}
	\renewcommand{\figurename}{FIG.|}
	\caption{\label{fig2}
        (a) Calculated absorption spectrum of hBN encapsulated monolayer MoS$_{2}$ with and (c) without electron-hole exchange interaction $U_{eh}$ for a carrier density of $n_s = 0.1 \cdot 10^{12} \SI{}{\per\centi\meter\squared}$.
        A phenomenological homogeneous broadening of $\SI{4}{\milli\electronvolt}$ (FWHM) has been used.
		(b),(d) Corresponding wave function contributions from the $T^{\text{inter}}_{S}$, $T^{\text{intra}}_{S}$ and $T^{inter}_{T}$ of the trion configurations to the absorption spectrum of hBN encapsulated monolayer MoS$_{2}$.
        (e) Calculated binding energies ($E- E_{X^{0}}$) of trion resonances, without $U_{eh}$, only with non-local $U_{eh}$ and with both local + non-local $U_{eh}$. The latter forms quantum superpositions by strongly admixing the trion eigenstates $T^{\text{intra}}_{S}$ and $T^{\text{inter}}_{T}$ shown in (a).
		(f) Binding energies of trions $T_1$, $T_2$ and $T_3$ as a function of the ratio between effective electron mass of upper and lower conduction band. Results are obtained using an effective mass model for the band structure around the $K$/$K^{\prime}$ points of the two lowest conduction (highest valence) bands.
		}
\end{figure}

In order to explain the threefold trion fine structure, we numerically solve a generalized three-particle Schrödinger equation to determine resonances in the optical absorption of hBN encapsulated monolayer MoS$_2$ at low carrier concentration ($n_s = 0.1 \cdot 10^{12} \SI{}{\per\centi\meter\squared}$)~\cite{Florian.2018}. The method used is combined with material-realistic band structures and bare, as well as screened Coulomb matrix elements on a G$_0$W$_0$ level (see SM). Results for the calculated optical absorption spectra are presented in Fig.~\ref{fig2}(a) and reveal $T_1$, $T_2$ and $T_3$ as individual peaks. Their experimentally observed PL counterparts can be identified in Fig.~\ref{fig1}(e). From our calculations, we can directly show that the observed features have distinct wave function contributions from the unperturbed trion configurations (see Fig.~\ref{fig2}(b)). For example, the lowest energy peak, $T_1$, contains contributions from $T^{\text{intra}}_{S}$ and $T^{\text{inter}}_{T}$, with the former dominating the total wave function. The $T_3$ resonance is similarly admixed with the dominant wave function contribution from the \textit{inter}valley triplet trion. Only $T_2$ is an eigenstate of the \textit{inter}valley singlet trion. Note that without electron-hole exchange interaction $U_{eh}$, the observed trion fine structure only shows two resonances (see Fig.~\ref{fig2}(c)) with the corresponding wave function contributions in Fig.~\ref{fig2}(d). 

Figure~\ref{fig2}(e) shows the evolution of the calculated trion binding energies ($E- E_{X^{0}}$) with local and non-local $U_{eh}$. Without $U_{eh}$, only two energetically distinct resonances are expected, which are conventionally labeled in the literature as the inter/intravalley trions. In contrast, including electron-hole exchange interaction predicts three resonances, as observed in our experiments. The absolute and relative energies of $T_1$, $T_2$ and $T_3$ are in excellent agreement with our experimental findings. Moreover, since electrons from both spin-orbit split conduction bands $c_1$ and $c_2$ contribute to singlet \textit{intra}valley trion and Coulomb exchange split \textit{inter}valley triplet trion, their corresponding binding energy sensitively depends on the ratio of the electron band masses $m_e^{c_1} / m_e^{c_2}$ (see Fig.~\ref{fig2}(f)). Hence, the trion fine structure contains additional information on the difference of the electron effective masses in the $c_1$ and $c_2$ conduction bands. The  $T_2$ trion remains unaffected since both spins are located in the same conduction band. 

A qualitative understanding of the trion fine structure can be obtained from a configuration model~\cite{Deilmann.2017,Torche.2019}. The homogeneous part of the equation of motion for the trion amplitude $T^{e_1e_2h_3}_{\mathbf k_1 \mathbf k_2 \mathbf k_3}$ constitutes a three-particle Hamiltonian whose eigenstates describe trions with total momentum $\mathbf Q$ (see SM for details). The Hamiltonian can be split into a part $H_0$ without electron-hole exchange and an exchange part $H_U$ according to the Coulomb matrix element.
Configurations are eigenstates of the three-particle Hamiltonian $H_0$ that contains the kinetic energies of two electrons and a hole as well as the direct Coulomb interaction. There are six optically bright trion configurations in the subspace of zero-momentum trions with a hole located in the highest valence band~\cite{Courtade.2017}. Due to time-reversal symmetry, the configurations are pairwise degenerate and are connected by changing $K$ into $K^{\prime}$ and flipping all spins. In Fig.~\ref{fig1}(c) the configurations are shown for $\mathbf Q=\mathbf K$. By adding electron-hole exchange to this picture, interaction between the configurations is introduced and leads to new eigenstates and -energies. It is these new eigenstates that are observed in our experiments, rather than the unmixed configurations. 

Throughout the remainder of the manuscript we label the most prominent trion feature $T_1$ as the "negative trion" $X^{-}$ to directly link it to other reports in the literature.

\begin{figure}
	\scalebox{\figurescale}{\includegraphics[width=1\linewidth]{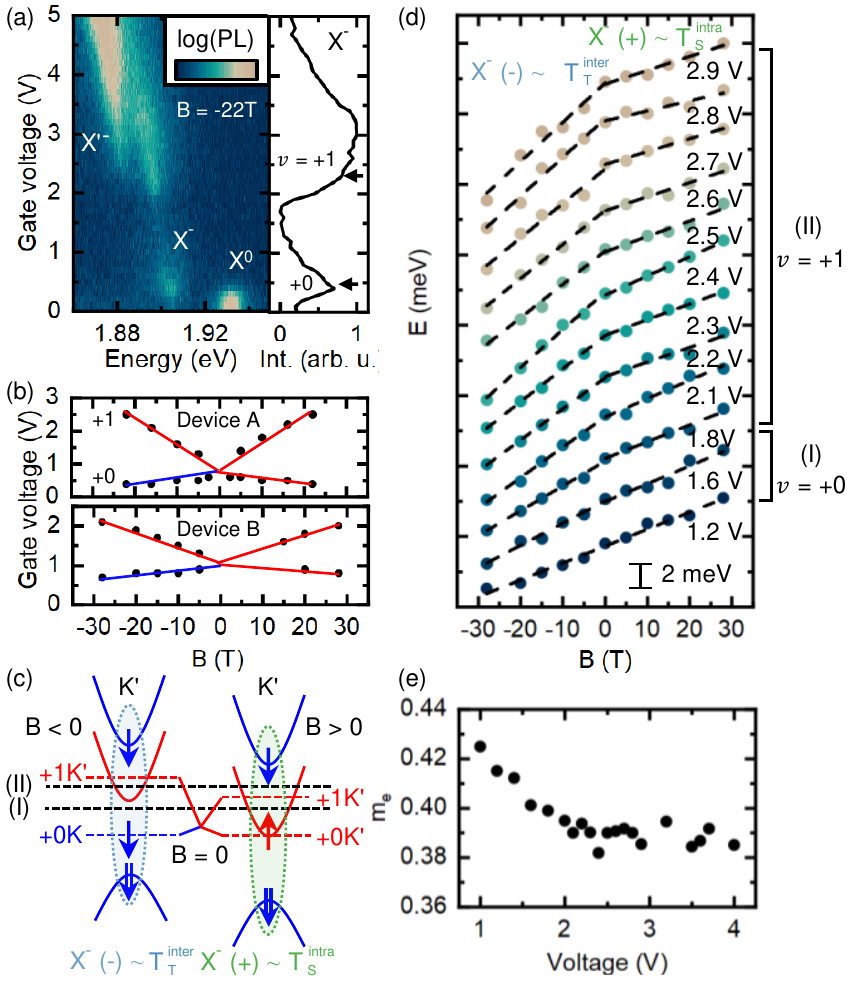}}
	\renewcommand{\figurename}{FIG.|}
	\caption{\label{fig3}
        (a) Left panel: Electron density dependent $\mu$-PL at $B = \SI{-22}{\tesla}$ of device A. Right panel: Corresponding integrated $X^-$ intensity. The $X^-$ reveals LL quantization as indicated by the dashed lines with filling factor $\nu = +0$ and $\nu = +1$. 
        (b) LL fan diagram of device A and B. Solid lines serve as a guide to the eye. The blue (red) color of the LLs corresponds to spin-$\uparrow$ (spin-$\downarrow$) of the electron for the magnetic field polarity due to time reversal symmetry breaking.
        (c) Band configuration including the relevant LLs in the lower conduction band of the \textit{\text{inter}}valley triplet trion $T_T^{\text{inter}}$ for negative (-) and \textit{\text{intra}}valley singlet trion $T_S^{\text{intra}}$ for positive (+) magnetic field. Regime I: Only the LL with filling factor $\nu = +0$ is filled in the $K$ ($K^{\prime}$) valley for negative (positive) magnetic field. Regime II: For negative field LL $\nu = +0$ is populated in $K$ while $\nu = +0$ and $\nu = +1$ are in $K^{\prime}$ for positive magnetic fields.
        (d) $X^-$ valley Zeeman shift in a second device B. The effective $g$-factor for negative magnetic field increases due to LL population effects.
        (e) Voltage dependent electron mass $m_e$.
		}
\end{figure}

We continue to probe the mixed character of the $X^{-}$ via magneto-optical experiments in high magnetic fields in device A and an additional, second, dual-gated device B. We focus our discussion on the $T_1$ resonance for which we obtain statistically reliable data for both devices. 
Figure~\ref{fig3}(a) shows a typical example of the PL recorded at $B = \SI{-22}{\tesla}$ as a function of gate voltage on device A. Clearly, the $X^{-}$-PL shows gate-voltage-dependent oscillations in its intensity (side panel of Fig.~\ref{fig3}(a)), which we can observe already at lower $B$ (see SM for additional data). The peaks of these quantum oscillations are attributed to half filling of LLs with filling factors $\nu=+0$ and $\nu=+1$~\cite{Liu.2020}. As detailed in the SM and plotted in Fig.~\ref{fig3}(b), the position of the peaks shift linearly with $B$, a general observation in both investigated devices. 

For low electron densities, we observe that the peak emission associated with the $\nu = +0$ LL shifts downwards in gate voltage (=energy) with increasing magnetic field strength. We explain this observation by the unique features of LLs in monolayer TMDCs ~\cite{Rose.2013}. Firstly, at finite magnetic field, all LLs in each respective $K$/$K^{\prime}$ valley are fully spin- and valley polarized. As a consequence, $X^-$ decays into a circularly polarized photon, the helicity of which depends on the $K$/$K^{\prime}$ valley, and an electron in a fully spin- and valley polarized Landau level (see Fig.~\ref{fig3}(c))~\cite{Rose.2013,Wang.2016,Liu.2020}. Uniquely, the $K$/$K^{\prime}$ position and, therefore, spin of the $\nu = +0$ LL depends on the polarity of the applied magnetic field, being either spin-$\downarrow$ at $B < 0$ or spin-$\uparrow$ at $B > 0$. In other words, both valley index and spin of the 0th Landau level in the conduction band flip depending on the polarity of $B$. Since we only detect $\sigma^{-}$ PL in our experiment (optical recombination in the $K^{\prime}$ valley), this requires the final state of $X^{-}$ to switch valleys depending on the B-field polarity. Because $X^{-}$ (trion resonance $T_1$) has wave function contributions from both $T_T^{\text{inter}}$ and $T_S^{\text{intra}}$ with an electron situated in the lower conduction band $c_1$ with either spin-$\downarrow$ or spin-$\uparrow$, the 'flavor' of this state is therefore determined by the polarity of the magnetic field, a direct consequence of the quantum superposition of its trion eigenstates at $B = \SI{0}{\tesla}$. The $X^{-}$ magneto-optical response directly reflects the properties of both trion configurations at zero magnetic field, while in our experiments, we selectively detect the $T^{\text{intra}}_{S}$ ($T^{\text{inter}}_{T}$) trion with a spin-$\uparrow$ electron (spin-$\downarrow$ electron) in $K^{\prime}$ ($K$) for $B > 0$ ($B < 0$). 

We continue to investigate the density dependent $X^{-}$ valley Zeeman shift $\Delta E_{VZ} = \frac{1}{2}g \mu_B B$ of $X^{-}$ for positive and negative magnetic field orientations ($g(s^+)$ and $g(s^-)$) at higher densities. The valley Zeeman shift directly encodes spin and orbital properties~\cite{Aivazian.2015,Srivastava.2015,MacNeill.2015,Li.2014,Stier.2016,Stier.2016b} and its gate voltage dependence is presented in Fig.~\ref{fig3}(d). When the electron density is low, such that only the $\nu = +0$ LL is occupied in $c_1$, the valley Zeeman shift for positive and negative magnetic fields are equal. However, for densities such that the $\nu = +1$ LL is occupied ($n_1$), a strong asymmetry emerges in positive and negative magnetic fields. A quantitative description of the trion valley Zeeman shift at fixed density has recently been reported for singlet and triplet trions in WSe$_2$~\cite{Lyons.2019}, consistent with our observations for MoS$_2$. The asymmetry of the $X^{-}$ valley Zeeman shift, $\Delta E_{VZ}^{X^-}(n,B)=1/2  (\tau_{z}g_{X^-}-\tau_{e}g_e)\mu_B B - g_l(n) \mu_B |B|$, originates in the density-dependent LL occupation, $g_l(n)$ of the final state, with the electron residing in a LL (for details, see SM). For an electron density above $n_1 \sim 3 \cdot 10^{12} \SI{}{\per\centi\meter\squared}$, the combination of spin, valley and cyclotron energy results in asymmetric LL dispersions in positive and negative magnetic fields for LLs with filling factors $\nu \geq +1$. As shown in Fig.~\ref{fig3}(d), this manifests itself as asymmetric slopes for positive and negative $B$-field directions. Since $\Delta E_{\hbar \omega}^+ + \Delta E_{\hbar \omega}^- = - (\hbar \omega_c - \hbar \omega_T)$, we determine the simple equation $ \Delta + \frac{1}{m_e}-\frac{1}{m_T} = 0$ with $\Delta = g_{ave}/2 = \frac{1}{2} (g(s^+) + g(s^-)) / 2 $ and the trion mass $m_T = 2m_e + m_h$, which describes the shift of a trion in a magnetic field (for details see SM). Considering a hole mass of $m_h = 0.6$~\cite{Eknapakul.2014}, we obtain an electron mass at low electron densities (regime I) of $m_e \sim 0.43$ that is in excellent agreement with literature (see Fig.~\ref{fig3}(e))~\cite{Kormnyos.2015}. For higher voltages (regime II), $m_e$ decreases due to the population of $c_1$ renormalizing the trion wave function due to many-body effects.





In summary, we have shown that trions in MoS$_2$ are quantum superpositions of \textit{inter}- and \textit{intra}valley spin states for the lowest electron densities close to the trion formation threshold. The observed exchange splitting and binding energies of the different trion species were shown to be strongly sensitive to electron-hole exchange effects. Moreover, pronounced non-uniformity in the Zeeman shift of the \textit{intra}valley trion revealed the importance of Landau level occupation dependent initial and final state energies. This fully accounts for recently observed variations in the exciton $g$-factor~\cite{Klein.2021}. Our results expand the current understanding of trion complexes in monolayer MoS$_2$ and show that their wave functions are strongly admixed, signatures of which are directly encoded in the evolution of valley Zeeman shifts. 


%
%
\section{Acknowledgements}
Supported by Deutsche Forschungsgemeinschaft (DFG) through the TUM International Graduate School of Science and Engineering (IGSSE) and the German Excellence Cluster-MCQST and e-conversion. We gratefully acknowledge financial support by the PhD program ExQM of the Elite Network of Bavaria. We also gratefully acknowledge financial support from the European Union’s Horizon 2020 research and innovation programme under grant agreement No. 820423 (S2QUIP) the German Federal Ministry of Education and Research via the funding program Photonics Research Germany (contracts number 13N14846) and the Bavarian Academy of Sciences and Humanities. M.F., A.S. and F.J. were supported by the Deutsche Forschungsgemeinschaft (DFG) within RTG 2247 and through a grant for CPU time at the HLRN (Berlin/G\"ottingen). J.K. and M.F. acknowledge support by the Alexander von Humboldt foundation. J.J.F and A.H. acknowledge support from the Technical University of Munich - Institute for Advanced Study, funded by the German Excellence Initiative and the European Union FP7 under grant agreement 291763 and the German Excellence Strategy Munich Center for Quantum Science and Technology (MCQST). Moreover, J.J.F. gratefully acknowledges the DFG for financial support via FI 947/8-1 and DI 2013/5-1 of SPP-2244. K.W. and T.T. acknowledge support from the Elemental Strategy Initiative conducted by the MEXT, Japan ,Grant Number JPMXP0112101001, JSPS KAKENHI Grant Number JP20H00354 and the CREST(JPMJCR15F3), JST. The work has been partially supported by the EC Graphene Flagship project, by the ANR projects ANR-17-CE24-0030 and ANR-19-CE09-0026. M.F., A.S. and F.J. thank Gunnar Schönhoff, Malte Rösner and Tim Wehling for providing material-realistic band structures and bare as well as screened Coulomb matrix elements.

\section{Author contributions}
J.K. and M.F. contributed equally to this work, J.K., A.Hö., A.H., M.P., C.F., J.J.F. and A.V.S. conceived and designed the experiments, A.Hö. and J.K. prepared the samples, K.W. and T.T. provided high-quality hBN bulk crystals, J.K., A.Hö., A.D., C.F. and A.V.S. performed the optical measurements, J.K. and A.V.S. analyzed the data, M.F., A.S. and F.J. computed the trion fine-structure, J.K., M.F., A.S., A.V.S and J.J.F. wrote the manuscript with input from all co-authors. \\

%



%
%

\bibliographystyle{naturemag}
\bibliography{full}

\end{document}


\title{Supplemental Material - Trions in MoS$_2$ are quantum superpositions of \textit{intra}- and \textit{inter}valley spin states}
%
\author{J.~Klein}\email{jpklein@mit.edu}
\affiliation{Walter Schottky Institut and Physik Department, Technische Universit\"at M\"unchen, Am Coulombwall 4, 85748 Garching, Germany}
\affiliation{Department of Materials Science and Engineering, Massachusetts Institute of Technology, Cambridge, Massachusetts 02139, USA}
\author{M.~Florian}
\affiliation{Institut für Theoretische Physik, Universität Bremen, P.O. Box 330 440, 28334 Bremen, Germany}
\affiliation{Department of Electrical Engineering and Computer Science, University of Michigan, Ann Arbor, MI, USA}
\author{A.~H\"otger}
\affiliation{Walter Schottky Institut and Physik Department, Technische Universit\"at M\"unchen, Am Coulombwall 4, 85748 Garching, Germany}
\author{A.~Steinhoff}
\affiliation{Institut für Theoretische Physik, Universität Bremen, P.O. Box 330 440, 28334 Bremen, Germany}
\author{A.~Delhomme}
\affiliation{Universit\'e Grenoble Alpes, INSA Toulouse, Univ. Toulouse Paul Sabatier, EMFL, CNRS, LNCMI, 38000 Grenoble, France.}
\author{T.~Taniguchi}
\affiliation{Research Center for Functional Materials, National Institute for Materials Science, 1-1 Namiki, Tsukuba 305-0044, Japan
}
\author{K.~Watanabe}
\affiliation{Research Center for Functional Materials, National Institute for Materials Science, 1-1 Namiki, Tsukuba 305-0044, Japan
}
\author{F.~Jahnke}
\affiliation{Institut für Theoretische Physik, Universität Bremen, P.O. Box 330 440, 28334 Bremen, Germany}
\author{A.~W.~Holleitner}
\affiliation{Walter Schottky Institut and Physik Department, Technische Universit\"at M\"unchen, Am Coulombwall 4, 85748 Garching, Germany}
\author{M.~Potemski}
\affiliation{Universit\'e Grenoble Alpes, INSA Toulouse, Univ. Toulouse Paul Sabatier, EMFL, CNRS, LNCMI, 38000 Grenoble, France.}
\author{C.~Faugeras}
\affiliation{Universit\'e Grenoble Alpes, INSA Toulouse, Univ. Toulouse Paul Sabatier, EMFL, CNRS, LNCMI, 38000 Grenoble, France.}
\author{A.~V.~Stier}\email{andreas.stier@wsi.tum.de}
\affiliation{Walter Schottky Institut and Physik Department, Technische Universit\"at M\"unchen, Am Coulombwall 4, 85748 Garching, Germany}
\author{J.~J.~Finley}\email{finley@wsi.tum.de}
\affiliation{Walter Schottky Institut and Physik Department, Technische Universit\"at M\"unchen, Am Coulombwall 4, 85748 Garching, Germany}
%

%
\date{\today}
%

%
\maketitle
%
%

\tableofcontents

\newpage

\section{Field-effect device for carrier density control in monolayer MoS$_{2}$}

%
\begin{figure}[ht]
\scalebox{\figurescale}{\includegraphics[width=0.6\linewidth]{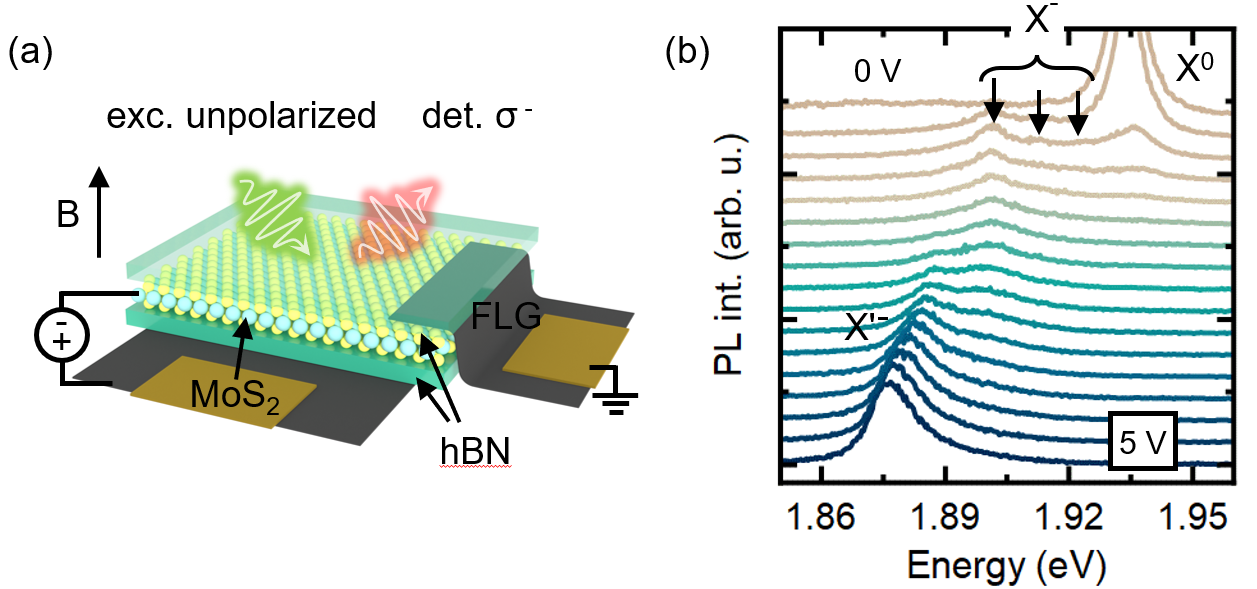}}
\renewcommand{\figurename}{SM Figure}
\caption{\label{SIfig1}
(a) Schematic illustration of the field-effect van der Waals devices investigated. The device $A$ is a single-gate device. Device $B$ is a dual-gate device with an additional few-layer graphite (FLG) gate on the top hBN~\cite{Klein.2021}. In the latter case, the carrier density is controlled with a top- and bottom-gate $V_{tg}$ and $V_{bg}$. (b) Waterfall plot of the gate voltage dependent PL on device $A$.
}
\end{figure}
%

We control the carrier density in the MoS$_{2}$ using a field-effect device geometry fabricated by the dry visco-elastic transfer method.~\cite{CastellanosGomez.2014} SM Figure~\ref{SIfig1}(a) shows a schematic illustration of the device used for our measurements. Monolayer MoS$_{2}$ is fully encapsulated in hBN to reduce inhomogeneous linewidth broadening of excitons.~\cite{Wierzbowski.2017} The hBN also serves as a gate dielectric. We apply a bias to the few-layer graphite with respect to the contact to the MoS$_2$ established by another few-layer graphite flake. We conduct measurements on two different devices, a device with one bottom gate and a dual-gate device~\cite{Klein.2021} with an additional graphite electrode situated on the top hBN. For the latter, we apply equal gate voltages with the same polarity to the gates $V_{bg}=V_{tg}$ with respect to the monolayer MoS$_{2}$ for controlling the carrier density. In both devices, we approximate the carrier density by using a plate capacitor model where the device capacitance is $C = \epsilon_0 \epsilon_{hBN}/d$ for the single-gate device with the dielectric constant of multilayer hBN $\epsilon_{hBN} = 2.5$~\cite{Dean.2010,Kim.2012,Hunt.2013,Laturia.2018} and a hBN layer thickness of $d \sim \SI{16}{\nano\meter}$ which is determined by atomic force microscopy (AFM). In the dual-gate device, top and bottom hBN thickness are both $d \sim \SI{14}{\nano\meter}$. Therefore, we can simply calculate the carrier density with the gate voltage through $n = C (V_{tg} + V_{bg}) /e = 2CV /e$.

For the experiment, the device is mounted on a xyz-piezo stack in a home-built probe stick system. The probe stick is dipped into a He cryostat with a sample lattice temperature for all experiments of $T = \SI{5}{\kelvin}$. We excite the monolayer with unpolarized light at $E = \SI{2.41}{\electronvolt}$ whereas only $\sigma^{-}$ light is detected to selectively probe the $K^{\prime}$ valley. Both, excitation and detection is fiber based. Detected light is filtered and then dispersed onto a charge-coupled device (CCD) for spectra acquisition. A typical waterfall plot of carrier density dependent PL spectra from device $A$ at zero magnetic field is shown in SM Fig.\ref{SIfig1}(b).

\newpage

\section{Theoretical calculation of trion spectra}

The linear absorption spectrum of MoS$_2$ was calculated by solving the semiconductor Bloch equations (SBE) for the microscopic interband polarizations $\psi^{he}_{\mathbf k} = \braket{a^h_{\mathbf k}a^{e}_{\mathbf k}}$, where the operators $a^{\lambda}_{\mathbf k}$ annihilate a carrier in band $\lambda$ with momentum $\mathbf k$. In the limit of vanishing excitation/doping density, the SBE become formally equivalent to the Bethe-Salpeter equation~\cite{rohlfing_electron-hole_2000} and read as follows in Fourier space
\begin{equation}
    \begin{split}
        &(\varepsilon^e_{\mathbf k} + \varepsilon^h_{\mathbf k} - \hbar\omega)\psi^{he}_{\mathbf k}(\omega) -  \\
        & \frac{1}{A}\sum_{\mathbf{k'}}\sum_{h'e'} \left( V^{eh'he'}_{\mathbf k,\mathbf k',\mathbf k,\mathbf k'} - U^{eh'e'h}_{\mathbf k,\mathbf k',\mathbf k',\mathbf k}\right)\psi^{h'e'}_{\mathbf k'}(\omega)  = (d^{he}_{\mathbf k})^*E(\omega)\, ,
    \end{split}
    \label{eq:SBE}
\end{equation}
Here, $\varepsilon^{e/h}_{\mathbf k}$ are the bandstructures in the valence / conduction bands,  $V^{\lambda_1\lambda_2\lambda_3\lambda_4}_{\mathbf k+\mathbf q,\mathbf k'-\mathbf q,\mathbf k',\mathbf k}$ are Coulomb matrix elements and $d^{he}_{\mathbf k}$ are interband dipole matrix elements that are projected into the polarization direction of the
electric field $E(t)$, assumed to be in the plane of the single-layer. Additionally, $A$ denotes the crystal area. Note that the electron-hole exchange interaction in Eq.~\eqref{eq:SBE} is described by unscreened Coulomb matrix elements $U$~\cite{sham_many-particle_1966,denisov_longitudinal_1973,Qiu.2015}, while $V$ denotes Coulomb matrix elements that are screened by carriers in occupied bands and the dielectric environment of the TMDC layer. The linear response of the material is given by the macroscopic susceptibility $\chi(\omega) = \frac{1}{A}\sum_{\mathbf k}\sum_{he}\left( \mathrm d^{he}_{\mathbf{k}}\psi^{he}_{\mathbf k} + c.c.\right)/E(\omega)$, which contains excitons as discrete resonances below a continuum of optical interband transitions.

%
%

As presented in Fig.~1 signatures of tightly bound trions appear in experimental spectra for moderate carrier densities.
To capture this effect, we extend the semiconductor Bloch equations presented in Eq.~\eqref{eq:SBE} and explicitly include correlation functions that describe three-particle states. For this purpose, we examine the equation of motion for the microscopic polarization for finite electron density, which reads
\begin{align}
    \begin{split}
        &(\varepsilon^e_{\mathbf k} + \varepsilon^h_{\mathbf k} - \hbar\omega)\Psi^{he}_{\mathbf k}(\omega) 
        -\\
        & \frac{1}{A}\sum_{\mathbf{k'}}\sum_{h'e'} \left(V^{eh'he'}_{\mathbf k,\mathbf k',\mathbf k,\mathbf k'} - U^{eh'e'h}_{\mathbf k,\mathbf k',\mathbf k',\mathbf k}\right)\Psi^{h'e'}_{\mathbf k'}(\omega) \\
        &+\frac{1}{A^2}\sum_{\mathbf{Q}\mathbf q}\sum_{e_2h_3e_4}\left( V^{e_2h_3he_4}_{\mathbf Q,\mathbf k-\mathbf q,\mathbf k,\mathbf Q-\mathbf q} - U^{e_2h_3e_4h}_{\mathbf Q,\mathbf k-\mathbf q,\mathbf Q-\mathbf q,\mathbf k} \right) \mathrm T_{ee_4h_3e_2}(\mathbf k, \mathbf Q-\mathbf q, \mathbf Q) \\
        &-\frac{1}{A^2}\sum_{\mathbf{Q}\mathbf q}\sum_{e_2e_3e_4}\left(V^{e_2ee_4e_3}_{\mathbf Q,\mathbf k,\mathbf k+\mathbf q,\mathbf Q-\mathbf q} - V^{e_2ee_3e_4}_{\mathbf Q,\mathbf k,\mathbf Q-\mathbf q,\mathbf k+\mathbf q}\right)\mathrm T_{e_4e_3he_2}(\mathbf k + \mathbf q, \mathbf Q-\mathbf q, \mathbf Q) \\
        &=(1 - f^e_{\mathbf{k}})(\mathrm d^{he}_{\mathbf k})^*E(\omega)\,.
    \end{split}
    \label{eq:eom_psi_trion}
\end{align}
Here, $f^\lambda_{\mathbf{k}}$ denotes the carrier population in the band $\lambda$. We consider the system to be in thermal quasi-equilibrium described by Fermi functions with given temperature and carrier density. The Coulomb interaction couples the polarization $\Psi^{he}$ to the trion amplitude
\begin{align}
    T_{e_1e_2h_3e_4}(\mathbf k_1, \mathbf k_2, \mathbf Q)=\braket{{a_{\mathbf{Q}}^{e_4}}^\dagger a_{-(\mathbf{k}_1+\mathbf{k}_2-\mathbf{Q})}^{h_3} a_{\mathbf{k}_2}^{e_2} a_{\mathbf{k}_1}^{e_1}}\,,
\end{align}
which is a four-operator expectation value and describes the correlated process of annihilating two electrons and one hole, leaving behind an electron with momentum $\mathbf Q$ in the conduction band and is linked to the optical response of an electron trion $X^-$. For moderate carrier density and linear optics a closed expression for the trion amplitudes is obtained~\cite{esser_theory_2001, Florian.2018}
\begin{align}
    \begin{split}
        &(\varepsilon^{e_1}_{\mathbf k_1} + \varepsilon^{e_2}_{\mathbf k_2} + \varepsilon^{h_3}_{\mathbf k_3}  - \varepsilon^{e_4}_{\mathbf Q} - \hbar\omega) T_{e_1e_2h_3e_4}(\mathbf k_1, \mathbf k_2, \mathbf Q) \\
        &-\frac{1}{A}\sum_{\mathbf q}\sum_{h_5,e_6}\left(V^{e_2h_5h_3e_6}_{\mathbf k_2,\mathbf k_3-\mathbf q,\mathbf k_3,\mathbf k_2-\mathbf q} - U^{e_2h_5e_6h_3}_{\mathbf k_2,\mathbf k_3-\mathbf q,\mathbf k_2-\mathbf q,\mathbf k_3}\right) T_{e_1e_6h_5e_4}(\mathbf k_1, \mathbf k_2- \mathbf q, \mathbf Q) \\
        &-\frac{1}{A}\sum_{\mathbf q}\sum_{h_5,e_6}\left( V^{e_1h_5h_3e_6}_{\mathbf k_1,\mathbf k_3-\mathbf q,\mathbf k_3,\mathbf k_1-\mathbf q} - U^{e_1h_5e_6h_3}_{\mathbf k_1,\mathbf k_3-\mathbf q,\mathbf k_1-\mathbf q,\mathbf k_3}\right) T_{e_6e_2h_5e_4}(\mathbf k_1- \mathbf q , \mathbf k_2, \mathbf Q) \\
        &+\frac{1}{A}\sum_{\mathbf q}\sum_{e_5,e_6}\left(V^{e_1e_2e_5e_6}_{\mathbf k_1,\mathbf k_2,\mathbf k_2+\mathbf q,\mathbf k_1-\mathbf q} - V^{e_1e_2e_6e_5}_{\mathbf k_1,\mathbf k_2,\mathbf k_1-\mathbf q,\mathbf k_2+\mathbf q}\right) T_{e_6e_5h_3e_4}(\mathbf k_1- \mathbf q , \mathbf k_2+\mathbf q, \mathbf Q) \\
&=f^{e_1}_{\mathbf{Q}}\left( \mathrm d^{he}_{\mathbf{k_2}} \delta_{\mathbf k_1,\mathbf Q} \delta_{e,e_1} - \mathrm d^{he}_{\mathbf{k_1}}\delta_{\mathbf k_2,\mathbf Q}  \right)E(\omega)\,,
    \end{split}
    \label{eq:eom_trion}
\end{align}
where the homogeneous part of these equations represents a generalized three-particle Schr\"odinger equation in reciprocal space. The oscillator strength of a transition between a trion state with total momentum $\mathbf Q$ and an electron state are directly proportional to the carrier population $f^{\lambda}_{\mathbf{Q}}$, which is apparent from the inhomogenous part of Eq.~\eqref{eq:eom_trion}. Thus, trions visible in optical spectra are hosted in the $K$ and $K^{\prime}$ valleys of the two lowest conduction (highest valence) bands to which we limit our single-particle basis.

\begin{figure}[!ht]
\scalebox{\figurescale}{\includegraphics[width=0.7\linewidth]{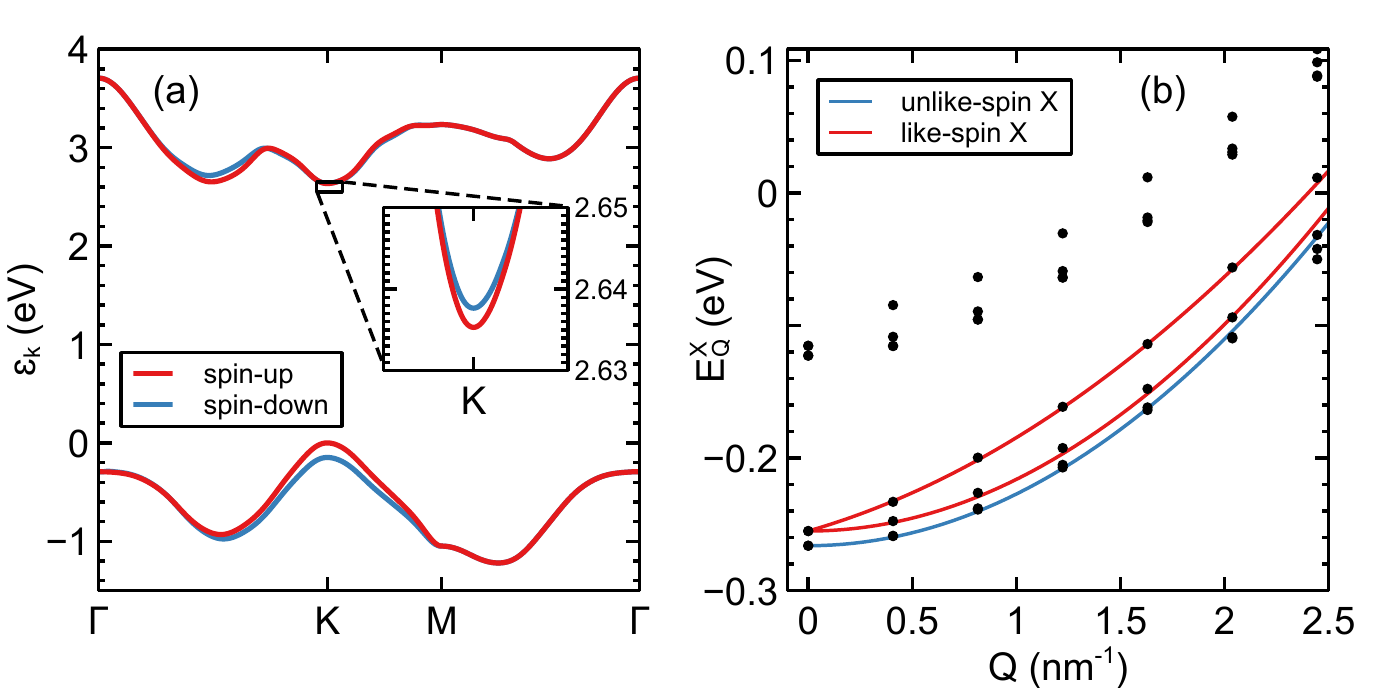}}
\renewcommand{\figurename}{SM Figure}
\caption{
Single-particle (a) and exciton (b) band structure of monolayer MoS$_2$. For the 1s-exciton dispersion, like- and unlike-spin bands are highlighted by red and blue solid lines, respectively, to demonstrate the dark-bright splitting induced by local electron-hole exchange. Due to this, the ordering of exciton bands with respect to spin configuration is different to the single-particle band structure so that the excitonic ground state is dark. Also, non-local exchange leads to a splitting of the bright-exciton dispersion into a parabolic lower branch and a nonanalytic upper branch.}
\label{SIfig2}
\end{figure}

We combine the above many-body theory of trions with ab-initio methods providing material-realistic band structures and bare as well as screened Coulomb matrix elements on a G$_0$W$_0$ level as input for the equations of motion (Eq. \eqref{eq:eom_psi_trion} and \eqref{eq:eom_trion}). More details on the G$_0$W$_0$ calculations are given in Ref.~\cite{Steinhoff.2014}. First- and second-order Rashba spin-orbit coupling are added subsequently in terms of a model Hamiltonian as introduced in Ref.~\cite{Liu.2013} to account for the spin-orbit splitting in the conduction- and the valence-band $K$ valleys. For the valence-band splitting, we obtain 148 meV, while the used spin-orbit coupling model yields a wrong sign for the conduction-band splitting, which has been discussed in Ref.~\cite{Liu.2013}. We therefore correct the second-order Hamiltonian (Eq.~(29) in Ref.~\cite{Liu.2013}) by adding $3$ meV to $\frac{3\lambda^2}{2(E_{+1}-E_0)}$ and subtracting $3$ meV from $\frac{3\lambda^2}{2(E_{-1}-E_0)}$. The resulting conduction-band splitting is $3$ meV, with the lowest conduction band having the same spin as the highest valence band in line with~\cite{Qiu.2015}. Solving the Bethe-Salpeter equation including electron-hole exchange interaction we find the unlike-spin KK excitons $11$ meV below the like-spin KK excitons which is comparable to the trend observed in Ref.~\cite{Qiu.2015}. For more details about the spin-orbit coupling, see Refs.~\cite{Liu.2013, Steinhoff.2014}. The band structure of monolayer MoS$_2$ including spin-orbit coupling is shown in SM~Figure~\ref{SIfig2}(a). 

Direct dipole transition matrix elements are calculated using a Peierls approximation~\cite{Steinhoff.2014}. Coulomb matrix elements for hBN encapsulated MoS$_2$ are obtained in two steps: Bare Coulomb matrix elements for a freestanding monolayer and the RPA dielectric function are calculated in a localized Wannier basis $\ket{\alpha}$ with dominant W-d-orbital character. Substrate effects are then included using the \textit{Wannier function continuum electrostatics} (WFCE) approach as described in Refs.~\cite{rosner_wannier_2015, Florian.2018}. The approach combines a continuum electrostatic model for the screening by the dielectric environment with a localized description of the Coulomb interaction. The actual parametrization is provided in Ref.~\cite{Florian.2018} and has been shown to provide reasonable agreement with experiments. Subsequently, the Coulomb matrix elements are transformed in Bloch-state representation
\begin{equation}
  \begin{split}
& U^{\lambda_1\lambda_2\lambda_3\lambda_4}_{\bk+\bq,\bkp-\bq,\bkp,\bk} = \\
        &\sum_{\alpha\beta\gamma\delta}
  \big[c_{\alpha}^{\lambda_1}(\bk+\bq)\big]^*\big[c_{\beta}^{\lambda_2}(\bkp-\bq)\big]^* \, c_{\gamma}^{\lambda_3}(\bkp) c_{\delta}^{\lambda_4}(\bk) \,\,
 U_{\bq}^{\alpha\beta\gamma\delta}\,.
    \label{eq:unitary_transform}
    \end{split} 
\end{equation}
where the coefficients $c_{\alpha}^{\lambda}(\bk)$ connect the localized and the Bloch basis as described in~\cite{Steinhoff.2014}. 
As the Coulomb interaction is spin-diagonal, only the electron-hole exchange terms couple trions with different spin combinations and are known to give rise to a fine structure splitting of the trion resonances~\cite{Yu.2014}. 
A proper description of electron-hole exchange requires matrix elements with density-density-like $U_{\bq}^{\alpha\beta\beta\alpha}$ and exchange-like contributions $U^{\alpha\beta\alpha\beta}$ in the local representation~\cite{Qiu.2015, steinhoff_biexciton_2018}. While the density-density-like matrix elements are momentum-dependent and thus nonlocal in real space, the exchange-like matrix elements are practically momentum-independent and correspond to local interaction processes between carriers within the same unit cell. For monolayer MoS$_2$ we find that the dominant exchange-like matrix elements amount to 0.378 eV (0.20 eV) per unit-cell area for interaction between $d_{m=0}$ and $d_{m=\pm2}$ (among $d_{m=\pm2}$) orbitals.
As discussed in \cite{Qiu.2015}, local electron-hole exchange leads to a splitting between dark (unlike-spin) and bright (like-spin) excitons, while nonlocal electron-hole exchange further modifies the dispersion of bright excitons. We demonstrate these effects by a numerical solution of the exciton Bethe-Salpeter equation, which represents the electron-hole subset of the trion equation (\ref{eq:eom_trion}):
\begin{align}
    \begin{split}
        &(\varepsilon^{e_1}_{\mathbf k_1}+ \varepsilon^{h_2}_{\mathbf k_2}) \Phi^{e_1h_2}_{\nu}(\mathbf k_1, \mathbf Q)-\frac{1}{A}\sum_{\mathbf q}\sum_{h_3,e_4}\left( V^{e_1h_3h_2e_4}_{\mathbf k_1,\mathbf k_2-\mathbf q,\mathbf k_2,\mathbf k_1-\mathbf q} - U^{e_1h_3e_4h_2}_{\mathbf k_1,\mathbf k_2-\mathbf q,\mathbf k_1-\mathbf q,\mathbf k_2}\right)
        \Phi^{e_4h_3}_{\nu}(\mathbf k_1- \mathbf q, \mathbf Q) \\
&=E^{\textrm{X}}_{\nu,\mathbf Q}\Phi^{e_1h_2}_{\nu}(\mathbf k_1, \mathbf Q)\,,
    \end{split}
    \label{eq:BSE}
\end{align}
with $\mathbf k_2=\mathbf k_1-\mathbf Q$. Diagonalization yields the exciton band structure including electron-hole exchange interaction shown in SI~Figure~\ref{SIfig2}(b). For the exchange-splitting between like- and unlike-spin excitons we obtain 11 meV.

To compute optical absorption spectra numerically, the coupled equations of motion (Eq.~\eqref{eq:eom_psi_trion} and \eqref{eq:eom_trion}) are solved in frequency space. Due to translational symmetry of the crystal, trions with different total momentum $\mathbf Q$ do not couple to each other, thus, we restrict the calculation to the main contribution at the $K$/$K^{\prime}$ point. Note that contributions from finite center-of-mass momentum give rise to an asymmetric line shape caused by electron recoil~\cite{esser_theory_2001}. The first Brillouin zone is sampled by a 84 $\times$ 84 Monkhorst-Pack mesh and a region around the $K$/$K’$ point defined by a radius of 3.5\,nm$^{-1}$. Due to the numerical discretization of momentum space, there is an uncertainty of binding energies of $\pm 0.5$ meV. For the matrix inversion problem we utilize an iterative Krylov-space method as contained in the PETSc toolkit~\cite{balay_petsc_2016}.

%
%

\newpage

\section{Landau levels in the negatively charged trion}

%
%
\begin{figure}[!ht]
\scalebox{\figurescale}{\includegraphics[width=1\linewidth]{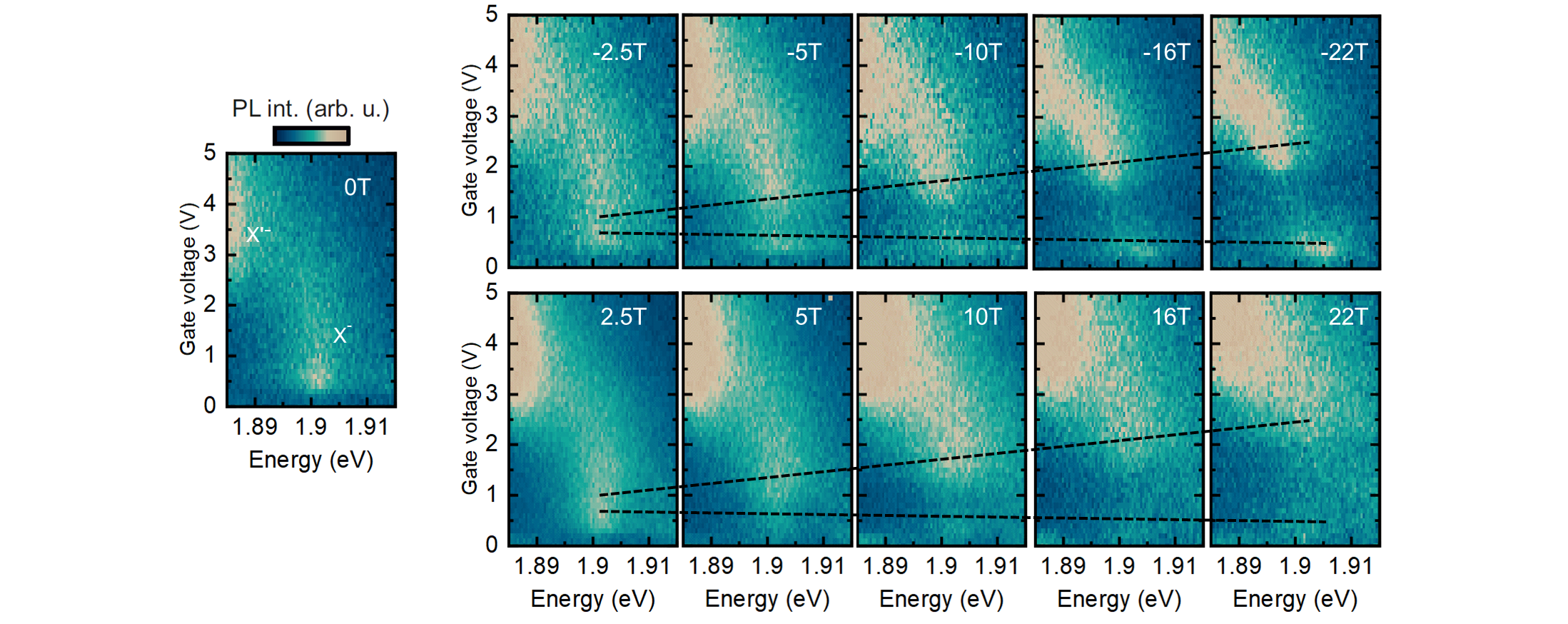}}
\renewcommand{\figurename}{SM Figure}
\caption{
Carrier density dependent low-temperature ($T = \SI{5}{\kelvin}$) $\mu$-PL of device $A$ for static magnetic fields ranging from $\SI{22}{\tesla}$ to $\SI{-22}{\tesla}$ of the \textit{intra}valley trion $X^-$. The dashed lines highlight Landau levels in the $X^-$ with filling factors of $\nu = +0$ and $\nu = +1$. Peaks in the integrated PL correspond to half filling of LLs. \cite{Liu.2020}
}
\label{SIfig7}
\end{figure}

\begin{figure}[!ht]
\scalebox{\figurescale}{\includegraphics[width=1\linewidth]{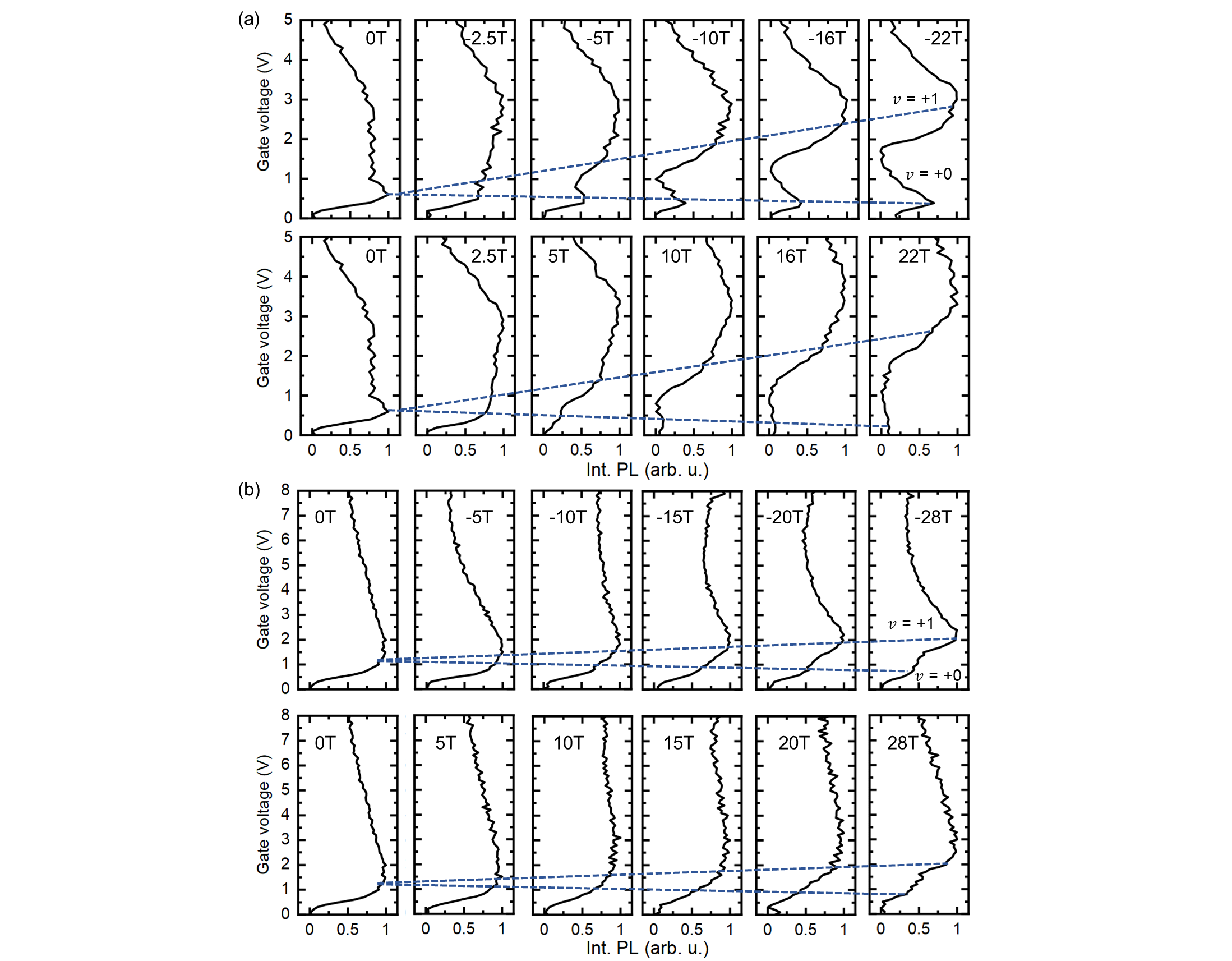}}
\renewcommand{\figurename}{SM Figure}
\caption{
(a) Spectrally integrated PL of the intravalley trion $X^-$ as a function of gate voltage for device $A$ and (b) device $B$. For increasing magnetic fields, LLs emerge with filling factors of $\nu = +0$ and $\nu = +1$ as indicated by the dashed lines.
}
\label{SIfig8}
\end{figure}

SM Figure~\ref{SIfig7} shows gate voltage dependent false color plots of the low-temperature ($T = \SI{5}{\kelvin}$) magneto-PL from device $A$. In particular, the intravalley trion $X^-$ is shown for magnetic fields ranging from $\SI{-22}{\tesla}$ to $\SI{22}{\tesla}$. Similar to the \textit{intra}valley trion in our device, Landau levels with filling factors $\nu = +0$ and $\nu = +1$ which period increases with the applied magnetic field. The quantum oscillations are analogous to Shubnikov-de-Haas oscillations in transport studies.~\cite{Pisoni.2018} 
This oscillation is clearly visible in the \textit{intra}valley trion $X^-$. SM Figure~\ref{SIfig8}(a) and (b) show the integrated PL intensity of the $X^-$ ($T_1$ resonance) as a function of the applied carrier density for static applied magnetic fields in device $A$ and device $B$. We find the same LL structure in both devices with filling factors of $\nu = +0$ and $\nu = +1$. The LLs vanish in both devices around the same gate voltage which is defined by the situation when the Fermi level touches the upper conduction band. Most importantly, the lowest LL with filling factor $\nu = +0$ shifts to lower gate voltages (carrier densities) with higher magnetic field.

\newpage

%
%

\section{Calculating the density dependent electron mass}

We now determine the effective electron mass $m_e$ based on the assumption that the hole mass is $m_h = 0.6$.~\cite{Eknapakul.2014} The Zeeman energy of the trion is given by~\cite{Lyons.2019}

\begin{equation}
    H_Z = \frac{1}{2} \bigg[\tau_1 g_{ve}^{vb} + s_{1}g_{s}^{vb}+\sum_{i=2}^{N=3}\bigg(\tau_ig_{ve}^{cb}+s_ig_s^{cb}\bigg)\bigg]\mu_B B
    \label{}
\end{equation}

where the index $i = 1$ is assigned to the bound hole and indices $i = 2, 3$ are assigned to the bound and excess electron of the trion. Here, $g_{ve}$ and $g_s$ are the valley and spin $g$-factors in the conduction and valence band, respectively. The valley and spin indices are $\tau_i = \pm 1$ ($K=+1$ and $K^{\prime}=-1$) and $s_i = \pm 1$ ($+1$ spin-$\uparrow$ and $-1$ spin-$\downarrow$). The change in the emitted photon energy due to radiative recombination of the negative trion in a magnetic field is~\cite{Lyons.2019}

\begin{equation}
\Delta E_{\hbar \omega} = \Delta E_Z - \Delta E_R + \Delta E_D
    \label{}
\end{equation}

with $\Delta E_Z$ the Zeeman shift of the electron-hole pair within the trion, $\Delta E_R$ the excess electron recoil shift and $\Delta E_D$ the diamagnetic shift. Since in the applied magnetic field range here, the diamagnetic shift is negligible,~\cite{Stier.2018} we omit this term in the following discussions. Unlike neutral exciton recombination, trion recombination cannot be a zero momentum process as the same momentum is imprinted onto the excess electron as recoil, detracting from the absorbed photon energy. The excess electron recoil depends on the temperature and the Fermi energy. The recoil cost is the energy difference in the trion and the free electron in the first LL. The change in energy of the excess electron after trion recombination in a magnetic field is~\cite{Lyons.2019}

\begin{equation}
    \Delta E_{e}^{\tau} = \frac{1}{2} \hbar \omega_c \bigg(1- \frac{m_e}{m_T}\bigg) + \frac{1}{2} \tau_e g_e \mu_B B
    \label{}
\end{equation}

with the cyclotron energy $\hbar \omega_c = \frac{e B}{m_e}$, the valley index $\tau_e$, the trion effective mass $m_T = \sum_{i = 1}^{N = 3} m_i$ and the bare electron $g$-factor $g_e$. Importantly, this recoil energy cost is equal in both valleys and simplifies to

\begin{equation}
   \Delta E_{e}^{\tau} = \frac{1}{2} \hbar \omega_c - \frac{1}{2} \hbar \omega_T + \frac{1}{2} \tau_e g_e \mu_B B
    \label{}
\end{equation}

where the first term describes the final state $\ket{f}$ of the trion, which is an electron in a LL, and the second term is the initial trion state $\ket{i}$. The difference is the recoil energy. The third term is the electron spin Zeeman energy that has no valley contribution. From these considerations, we can write the energy Zeeman shift for positive and negative magnetic field as measured in our experiment. For positive magnetic field we obtain

\begin{equation}
    \Delta E_{\hbar \omega}^+ = E_{Z} - \bigg[\frac{1}{2} (\hbar \omega_c - \hbar \omega_T) + \frac{1}{2} g_e \mu_B B\bigg]~,
    \label{}
\end{equation}

while for negative field 

\begin{equation}
    \Delta E_{\hbar \omega}^- = - E_{Z} - \bigg[\frac{1}{2} (\hbar \omega_c - \hbar \omega_T) - \frac{1}{2} g_e \mu_B B\bigg]~.
    \label{}
\end{equation}

The sum of the above Zeeman shifts for positive and negative field is 

\begin{equation}
    \Delta E_{\hbar \omega}^+ + \Delta E_{\hbar \omega}^- = - (\hbar \omega_c - \hbar \omega_T)~,
\end{equation}

which is proportional to the difference between the cyclotron energy of an electron in a LL and the trion LL. This can be expressed as

\begin{equation}
    \Delta E_{\hbar \omega}^+ + \Delta E_{\hbar \omega}^- = - \hbar e B\bigg(\frac{1}{m_e}-\frac{1}{m_T}\bigg)~,
\end{equation}

with $m_T = 2 \cdot m_e \cdot m_0 + m_h \cdot m_0$. From our experiment we directly determine the Zeeman shift and corresponding $g$-factor of the negatively charged trion as a function of the gate voltage (carrier concentration). Hence, we can relate the measured $g$-factor for positive and negative magnetic field $g(s^+)$ and $g(s^-)$ to the energy difference of initial and final state of the trion, and therefore to the electron mass taking into account a constant hole mass of $m_h = 0.6$ from Ref~\cite{Eknapakul.2014}. We obtain

\begin{equation}
   \frac{1}{2} g(s^+) \mu_B B + \frac{1}{2} g(s^-) \mu_B B = - \hbar e B \cdot \bigg(\frac{1}{m_e}-\frac{1}{m_T}\bigg)~,
\end{equation}

which we can rewrite to

\begin{equation}
   \frac{1}{2} \cdot g_{ave} = - \bigg(\frac{1}{m_e}-\frac{1}{m_T}\bigg)~,
\end{equation}

with $g_{ave} = 1/2 \cdot g(s^+) + g(s^-)$ and further simplify to

\begin{equation}
   \Delta + \frac{1}{m_e}-\frac{1}{m_T} = 0~,
\end{equation}

with $\frac{\mu_B}{\hbar e} = \frac{1}{2 m_e}$ and $\Delta = \frac{g_{ave}}{2}$. This equation describes the shift of a trion in a magnetic field. We can solve this equation for the electron mass resulting in

\begin{equation}
   m_{e(1,2)} = \frac{-1-\Delta m_h \pm \sqrt{1-6\Delta m_h + \Delta^2 m_h^2}}{4 \Delta}~,
\end{equation}

with $m_T = 2 \cdot m_e + m_h$. Here, $m_h$ is the hole mass in the upper valance band. In our definition, $m_e$ is negative which is the case for $m_{e(2)}$.






%
%


\newpage

%
%
%
%
\bibliographystyle{apsrev}
\bibliography{full}